\documentclass[aps,prl,twocolumn,groupedaddress,floats,showpacs]{revtex4}
\usepackage{latexsym}
\usepackage{dcolumn}
\usepackage[dvips]{graphicx}
\usepackage{amssymb}
\usepackage{graphics}
\usepackage{amsmath}
\usepackage{epsf}

\newcommand{\vk}{{\bf k}}

\newcommand{\ceq}[1] {(\ref{#1})}
\newcommand{\rr}     {{\bf r}}
\newcommand{\mn}     {m}

\begin{document}

%\twocolumn[
%\hsize\textwidth\columnwidth\hsize\csname@twocolumnfalse\endcsname
%\draft

\title{Theory of thermopower in 2D graphene}
\author{E. H. Hwang, E. Rossi, and S. Das Sarma}
\address{Condensed Matter Theory Center, Department of Physics,
  University of Maryland, College Park, 
Maryland  20742-4111 } 
\date{\today}

\begin{abstract}
Motivated by recent experiments \cite{Kim,Wei,Ong} 
we calculate the
thermopower of 
graphene incorporating the energy dependence of various transport scattering
times. 
%We consider the scattering by charged impurities, neutral
%white noise disorder, and acoustic phonons. 
We find that 
scattering by screened charged impurities gives a reasonable
explanation for the measured
thermopower. The calculated thermopower behaves as
$1/\sqrt{n}$ at high densities, but saturates at low densities.
We also find that the thermopower scales with normalized
temperature $T/T_F$ 
and does not depend on the impurity densities, but strongly depends on the fine
structure 
constant $r_s$ and the location of the impurities. 
We discuss the deviation from the Mott formula in graphene
thermopower, and use an effective medium theory to calculate
thermopower at low carrier density regimes where electron-hole
puddles dominate. 

\pacs{81.05.Uw; 73.61.Ey, 73.50.Jt, 71.30.+h}

\end{abstract}
\vspace{0.5in}
%]

\maketitle

Thermopower has been used as a powerful tool to probe transport mechanisms
in metals and semiconductors. Often the measurement of resistivity 
(or conductivity) is inadequate in distinguishing among different 
scattering mechanisms and the thermopower can then be used as a
sensitive probe of transport properties
since it provides complementary information to resistivity. 
In this Letter we develop a theory for the thermopower of
graphene with a goal toward elucidating the comparative importance of
various scattering mechanisms in graphene transport properties.

Recently, the thermoelectric properties of graphene have
attracted experimental attention  \cite{Kim,Wei,Ong,Balandin}.
Experimentally \cite{Kim,Wei,Ong}
the expected change of sign in the thermopower is found across the
charge neutral point  
as the majority carriers change from electrons to holes.
Away from the charge neutral region
the density dependence of thermopower behaves as $1/\sqrt{n}$, and 
exhibits a linear temperature dependence in agreement with the
semiclassical Mott
formula \cite{Mott}. 
As the temperature increases, a deviation from Mott formula is
reported\cite{Kim,Wei}. 
Existing theoretical works on graphene thermopower 
either use an impurity band model 
or consider the low temperature Mott limit \cite{thermo_th}. 
In this paper we present a calculation of the thermopower of graphene
taking into account the energy dependent scattering time for various 
scattering mechanisms. Understanding 
thermopower requires a thorough knowledge of
the details of
energy dependence of transport scattering
times \cite{Ashcroft,Shayegan,Lyo}. In metals the Mott formula 
is widely used because the Fermi temperature is very high (i.e. $T \ll
T_F$) 
and the scattering time is essentially energy independent leading to a
simple linear-in temperature form for thermopower which is
proportional to the energy derivative of the conductivity evaluated at
the Fermi energy.
Mott formula, derived mathematically through the Sommerfeld
expansion, is only valid at very low temperatures, $T/T_F \ll 1$.

We show that scattering by random charged impurity centers,
which is the main scattering mechanism limiting graphene
conductivity,\cite{Hwang_PRL} also dominates its thermopower. 
We show that the temperature dependent screening effects 
\cite{Hwang_sc} must be 
included in the theory to get quantitative agreement with
existing experimental data. 
We find the effects of short-range scattering and phonons to be
negligible 
in experimental temperature range ($T < 300K$), allowing us to ignore
phonon drag contribution.
We also find that
the calculated thermopower scales with $T/T_F$ and manifests
no impurity density ($n_i$) dependence, but depends strongly on
the impurity location and the
dielectric constant of the substrate (or equivalently the fine
structure constant of 
graphene).  The experimentally observed 
asymmetry between electron and hole thermopower is explained by the
asymmetry in the charged impurity configuration in the presence of the
gate voltage. We explain the
experimentally observed  sign change near charge neutral point (Dirac point)
with a simple two component model, which we explicitly verify using an
effective medium theory calculation taking into account the
inhomogeneous puddle formation. \cite{Hwang_PRL}

The ratio of the measured voltage to the temperature gradient
applied across the sample is known as the Seebeck coefficient  (or the
thermopower)  and is given by $Q =  \nabla V/\nabla T$, 
where $\nabla V$ is the
potential difference and $\nabla T$ the temperature difference between two
points of the sample \cite{Ashcroft}.
%Within the relaxation time approximation the Fermi distribution
%function in the presence of a uniform static electric field and
%temperature gradient is given by \cite{Hwang_PRL}
%\begin{equation}
%f_{\vk} = f_{\vk}^0 - \tau(\epsilon_{\vk})
%\left (
%  \frac{\partial f_{\vk}}{\partial \epsilon_{\vk}} \right ) {\bf
%  v}_{\vk} \left [ -e {\bf 
%      E} + \frac{\epsilon_{\vk}-\mu}{T}(-\nabla T) \right ],
%\end{equation} 
%where $f_{\vk}^0$ is the equilibrium Fermi distribution function, $\tau$ 
%the relaxation time,  and $\mu$ the chemical potential.
In linear response approximation
for the electrical current density, ${\bf j}$, and thermal current
density, ${\bf j}_Q$, we have:
${\bf j}  =  L^{11} {\bf E} + L^{12}(-\nabla T)$, 
${\bf j}_Q =  L^{21} {\bf E} + L^{22} (-\nabla T)$,
where $L^{ij}$ is defined in terms of the $I^{(\alpha)}$, i.e.,
$L^{11}  =  I^{(0)}$, 
$L^{12}  =  -\frac{1}{eT} I^{(1)}$,
$L^{21}  =  -\frac{1}{e} I^{(1)}$, and
$L^{22}  =  \frac{1}{e^2T} I^{(2)}$.
Here, $I^{(\alpha)}$ is given by 
\begin{eqnarray}
I^{(\alpha)} & = & e^2 g \sum_{\vk} \tau(\epsilon_{\vk}) {\bf v}_{\vk}
{\bf v}_{\vk}  \left [ \epsilon_{\vk} - \mu \right ]^{\alpha} \left
  (- \frac{\partial f_{\vk}^0}{\partial \epsilon_{\vk}} \right )
\nonumber \\
& = & \int d\varepsilon \left ( \varepsilon - \mu \right )^{\alpha} \left
  (- \frac{\partial f^0(\varepsilon)}{\partial \varepsilon} \right )
\sigma(\varepsilon),
\end{eqnarray}
where $f_{\vk}^0$ is the equilibrium Fermi distribution function, $\tau$ 
the relaxation time,  and $\mu$ the chemical potential,
$g=g_sg_v$ is the total degeneracy ($g_s=2$, $g_v=2$ being the spin
and valley degeneracies, respectively), and
$\sigma(\varepsilon)$ is the energy dependent conductivity of
graphene given by 
$\sigma(\varepsilon) = {e^2 v_F^2}D(\varepsilon)
\tau(\varepsilon)/2$,
where $v_F$ is the Fermi velocity and $D(\varepsilon) = g
|\varepsilon|/(2\pi \hbar^2v_F^2)$ the density of states.
From the definition of the thermopower
we have $Q = \frac{L^{12}}{L^{11}}$,
and, $\sigma=L^{11}$.

\begin{figure}
\bigskip
\epsfxsize=.7\hsize
\hspace{0.0\hsize}
\epsffile{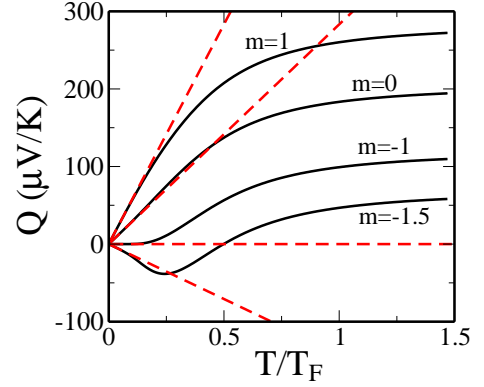}
\caption{\label{fig0}
%ER
%Calculated 
%
Hole thermopower for different energy dependent scattering
times, $\tau \propto \varepsilon^\mn$. For $\mn < - 1$ the low temperature
thermopower becomes negative. Dashed lines show the Mott formula for 
the corresponding scattering times. Electron results are the same with
an overall negative sign.
}
\end{figure}

Before we calculate the details of the thermopower for different 
scattering mechanisms we first consider the low temperature and high 
temperature behavior of $Q(T)$.
At low temperatures ($T\ll T_F$, where $T_F = E_F/k_B$) we can
express $I^{(\alpha)}$ as 
\begin{equation}
I^{(\alpha)}  =  \frac{1}{4\beta^{\alpha}} \int^{\infty}_{-\infty} dx
\frac{x^{\alpha}}{\tanh^2({x}/{2})} \left [ \sigma(\mu) + \frac{x}{\beta}
\left .  \frac{\partial \sigma(\varepsilon)}{\partial \varepsilon}
\right |_{\varepsilon = \mu} \right ]
\end{equation}
where $\beta = 1/k_BT$. Thus we have the well-known Mott formula
\cite{Mott} of
thermopower at low temperatures, i.e.
\begin{equation}
Q = - \frac{\pi^2}{3e} \frac{T}{\sigma(\mu)} \left . \frac{\partial
  \sigma(\varepsilon)}{ \partial \varepsilon} \right |_{\varepsilon =\mu}.
\label{Q_low1}
\end{equation}
If the energy dependence of the relaxation time is unimportant the
sign of the thermopower is determined by whether the carriers are
electrons or holes.
Assuming the energy dependent scattering time to be
$\tau \propto \varepsilon^\mn$ we have the thermopower at low
temperatures
\begin{equation}
Q = - \frac{\pi^2}{3e} \frac{k_B T}{T_F}(\mn+1).
\label{Q_low2}
\end{equation}
We note that in general the exponent `$m$' has weak temperature and density
dependence since $\tau$ behaves only as an effective power law in
energy. Eq. (\ref{Q_low2}) indicates that the
thermopower can change sign if $\mn<-1$. 
At high temperatures ($T \gg T_F$) we can express the $I^{(\alpha)}$
with an energy dependent scattering time $\tau = \tau_0 \varepsilon^{\mn}$ as
\begin{equation}
I^{(\alpha)} \approx {E_F^{\alpha}} \left ( \frac{T}{T_F} \right
)^{\alpha+\mn+1} \left [1- \frac{1}{2^{\alpha + \mn}} \right ]
\Gamma(\alpha + \mn+2) \zeta(\alpha + \mn + 1),
\end{equation}
where $\Gamma(z)$ and $\zeta(z)$ are the gamma function and Riemann's
zeta function, respectively, 
%ER
from which we find:
%Then we have the thermopower at high
%temperatures
\begin{equation}
Q \approx \frac{k_B}{e} \frac{(\mn+2)}{2} \frac{(2^{\mn+1}-1)}{(2^\mn-1)}
  \frac{\zeta(\mn+2)}{\zeta(\mn+1)}.
\end{equation}
%ER
At high temperatures $Q$ in graphene
approaches a limiting value.
%
%The high temperature thermopower of graphene approaches a limiting
%value.
In Fig.~\ref{fig0} we show the calculated graphene thermopower for different
scattering exponents $\mn$ ($\tau \propto \varepsilon^\mn$) as a
function of 
%ER
$T/T_F$.
%temperature. 
As shown in Fig.~\ref{fig0} the dashed lines representing Mott formula agree 
with the full calculations for $T \alt 0.2 T_F$. 
%If the Fermi energy 
%is very high the
%Mott formula is a good approximation for measured thermopower.
In addition the calculated thermopower scales as a function of the 
normalized temperature ($T/T_F$).

%%% Figure template for latex

Now we calculate the thermopower in the presence of various physical
scattering mechanisms.
For both neutral white noise short-range disorder and acoustic phonon
\cite{Hwang_ph} scattering we can express
the scattering time 
%in the low temperature limit 
as $\tau(\epsilon) = {\tau_1}/{\epsilon}$,
and we have 
%for $t = T/T_F \ll 1$
\begin{eqnarray}
I^{(0)} &=& \sigma_1 \frac{1}{1+e^{-\beta \mu}}, \\
I^{(1)} &=& \sigma_1   \left [ k_B T \ln[1+ e^{-\beta \mu}] +
  \frac{\mu}{1+ e^{\beta \mu}} \right ].
\end{eqnarray}
Then the thermopower becomes
\begin{equation}
Q = -\frac{1}{e}\left [ \frac{\mu}{T}e^{-\beta \mu} + O(e^{-\beta
    \mu}) \right ]. 
\end{equation}
The thermopower contributions from both neutral scatterers and
acoustic phonons  
are exponentially suppressed in the low temperature limit, and can be
ignored.  
For unscreened charged impurities 
we have very simple energy dependent scattering time \cite{Hwang_sc}
$\tau(\epsilon) = \tau_0 \epsilon$.
Then
we have the following integrals for $t \ll 1$
\begin{eqnarray}
I^{(0)} & = & \sigma_0\left [ 1 + O(e^{-\beta \mu}) \right ], \\
I^{(1)} & = & \sigma_0 E_F  \left [  \frac{2\pi^2}{3} t^2 + O(e^{-\beta \mu})
\right ],
\end{eqnarray}
leading to the thermopower 
\begin{equation}
Q = -\frac{2\pi^2}{3e}\frac{k_B^2 T}{E_F} \left [1+ O(e^{-\beta \mu}) \right ].
\label{q_coul}
\end{equation}
The linear relation of thermopower with temperature 
(or Mott formula) holds to relatively high temperatures in graphene
for unscreened Coulomb impurities.

As has been demonstrated theoretically  
and experimentally 
the dominant transport mechanism in graphene is the screened Coulomb scattering
from charged impurities.
The result of Eq.~(\ref{q_coul}) for unscreened Coulomb scattering is
much higher than the thermopower observed in experiments
\cite{Kim,Wei,Ong} and cannot
explain the behavior of $Q$ close to the Dirac point. An accurate
quantitative agreement between theory and experiment can only be
achieved by taking into account the screening of the charged
impurities and the strong spatial inhomogeneity that these impurities
induce close to the Dirac point.
For the screened charged impurity scattering, the energy dependent
scattering time  
$\tau(\varepsilon_k)$ is given \cite{Hwang_PRL}
\begin{eqnarray}
\label{eq:scattime}
\frac{1}{\tau(\epsilon_k)} & = & \frac{\pi n_i}{\hbar}
\int\frac{d^2k'}{(2\pi)^2}
\left |\frac{v_i(q)}{\epsilon(q,T)}\right |^2 \delta\left (
  \epsilon_{\bf k} - \epsilon_{\bf k'} \right ) \nonumber \\
& & \times (1-\cos\theta) (1 + \cos\theta),
\label{ttk}
\end{eqnarray}
where $\theta$ is the scattering angle,
$v_i(q)=2\pi e^2 \exp(-q d)/(\kappa q)$ is the Fourier transform 
of the 2D Coulomb potential in an effective background lattice
dielectric constant $\kappa$ and $d$ is the location of charged impurity
measured from graphene surface. 
In Eq.~(\ref{ttk}),
$\varepsilon(q)\equiv \varepsilon(q,T)$ is the 2D finite temperature
static RPA dielectric (screening) function appropriate for 
graphene\cite{Hwang_sc}, given by $\varepsilon(q,T) = 1 + v_c(q)
\Pi(q,T)$, where $\Pi(q,T)$ is the graphene irreducible
finite-temperature polarizability function and $v_c(q)$ is the Coulomb
interaction. 
There is an important  direct $T$ dependence of thermopower, not
captured in the Mott formula, arising from
the temperature dependent screening.~\cite{Hwang_sc} 
The temperature dependent conductivity due to screening effects 
decreases quadratically at low temperatures \cite{Hwang_sc}. This
mechanism produces a
thermopower quadratic in temperature rather than linear as in
the simple Mott formula.
Thus we predict a nonlinear quadratic temperature correction in the
graphene thermopower compared with the linear Mott formula.

%%% Figure template for latex
\begin{figure}
\bigskip
\epsfxsize=1.\hsize
\hspace{0.0\hsize}
\epsffile{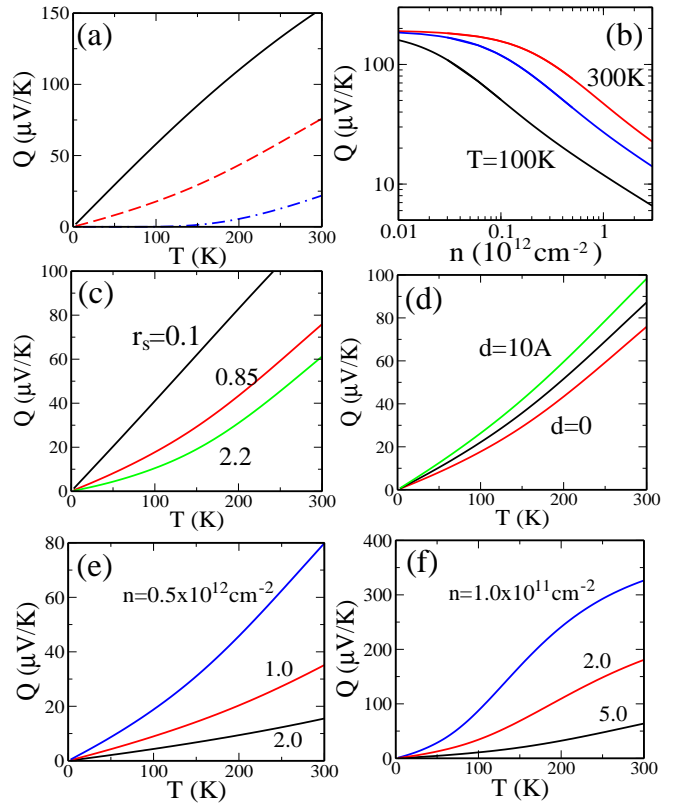}
\caption{\label{fig.1}
%ER
%(a) Calculated thermopower as a function of  temperature 
%for different scattering mechanisms. Solid, 
%dashed, and dot-dashed line represent unscreened Coulomb, screened Coulomb, 
%and neutral scatterers, respectively.
%(b) Calculated thermopower for the screened charged impurity scattering
%as a function of density for different temperatures. 
%Calculated thermopower (c) for different $r_s=0.1$, 0.85, 2.2 
%with $d=0$ and (d) for different
%$d=0$, 5, 10 \AA \;\;  with $r_s=0.85$.
%In (e) and (f), the calculated thermopower due to the screened charged
%impurities for different densities  is shown. In (e) we use
%parameters corresponding to graphene on SiO$_2$ with mobility
%$\mu = 10^4$ cm$^2$/Vs, and in (f) to suspended graphene with mobility
%$\mu =2\times 10^5$ cm$^2$/Vs.  
%
(a) $Q$ as a function of  temperature 
for different scattering mechanisms. Solid, 
dashed, and dot-dashed line represent unscreened Coulomb, screened Coulomb, 
and neutral scatterers, respectively.
(b) $Q$ for the screened charged impurity scattering
as a function of density for different temperatures. 
(c) $Q$ for different $r_s=0.1$, 0.85, 2.2 
with $d=0$ and (d) for different
$d=0$, 5, 10 \AA \;\;  with $r_s=0.85$.
In (e) and (f), $Q$ in the presence of screened charged
impurities for different densities  is shown. In (e) we use
parameters corresponding to graphene on SiO$_2$ with mobility
$\mu = 10^4$ cm$^2$/Vs, and in (f) to suspended graphene with mobility
$\mu =2\times 10^5$ cm$^2$/Vs.  
}
\end{figure}

In Fig.~\ref{fig.1}(a) we show the 
calculated thermopower of holes in graphene due to
different scattering mechanisms. 
The thermopower due to
screened charged impurity is about
half of that due to the unscreened charged impurity, and 
increases in a concave manner due to temperature and energy dependent
screening.
On the other hand, the
thermopower due to neutral scatterers is exponentially suppressed in the
low temperature regime.
Fig.~\ref{fig.1}(b) 
shows the calculated thermopower for the screened charged impurity scattering
as a function of density for different temperatures.
As we expect the density dependence 
shows $1/\sqrt{n}$ behavior at high densities. But this
power law behavior breaks down and
saturates at low densities. The saturation value ($Q_s$) 
does not depend on 
temperature. The theoretical $Q_s$ is just a function of interaction
(fine structure) parameter $r_s=e^2/\kappa \hbar v_F$ 
and the location of impurities $d$.
In Fig.~\ref{fig.1} (c) and (d) we show the $r_s$ and $d$ dependence of the
thermopower. In general the thermopower increases when the 
substrate dielectric constant ($\kappa$) increases or the charged
impurities move away.
Thus, we predict that the thermopower of suspended graphene
will decrease compared with the thermopower of graphene on a substrate
for the same densities because of the reduction of the dielectric constant. 
In Fig~\ref{fig.1} (e) and (f) we show the calculated thermopower for
systems with different mobilities due to screened charged impurity
scattering for graphene on SiO$_2$ substrate and for suspended graphene.

\begin{figure}
 \bigskip
 \epsfxsize=.80\hsize
 \hspace{0.0\hsize}
 \epsffile{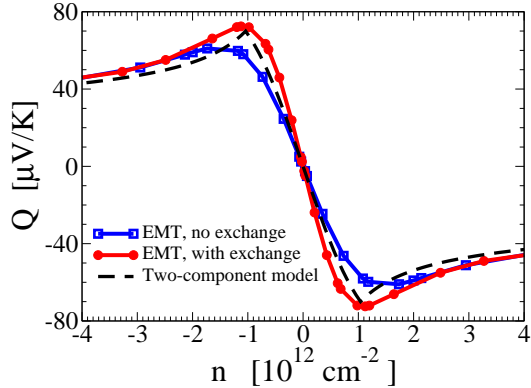}
 \caption{\label{fig.2}
 $Q$ due to the screened charge impurities as a function of density
 close to the Dirac point for  
 $T=300$~K, $n_{i}=10^{12}{\rm cm}^{-2}$, $r_s=0.8$ and  $d=1$~nm,
 obtained using the two-component model and the EMT with and without
 exchange energy.
 }
\end{figure}

In recent experiments \cite{Kim,Wei,Ong} 
it has been observed
that close to the Dirac point $Q$ does not follow the $1/\sqrt{n}$ scaling
predicted by the Mott formula.
The reason for this deviation is that close to the Dirac point
the quenched disorder induces strong density fluctuations that
break up the density landscape in electron-hole puddles 
\cite{Hwang_PRL, yacobi,rossi2008a,rossi2008b}.
To account for the main features of the thermopower close to the Dirac point
we can use a simple two component model near to the charge neutral
regime in which the electron density, $n_e$
and the hole density $n_h$ depend on the doping $n$ according
to the phenomenological equations: 
$ n_e =  (n_{\rm rms} + n)/2$,
$ n_h =  (n_{\rm rms} -n)/2$ for $|n| \le n_{\rm rms}$,
where $n_{\rm rms}$ is the root mean square of the density fluctuations.
We assume 
$n_{\rm rms}\sim n_{i}$ \cite{Hwang_PRL,rossi2008a}.
In the two component model the thermopower becomes
$Q=(L_e^{12}+L_h^{12})/(L_e^{11}+L_h^{11})$ 
so that, for unscreened charged impurities, we find:
\begin{equation}
 \label{eq:two-comp}
 Q = -\frac{k_B^2}{e} \frac{2\pi^2}{3}\frac{T}{v_F\sqrt{\pi}} \left [
  \frac{ \sqrt{n_e} - \sqrt{n_h}}{ n_e + n_h} \right ].
\end{equation}
Thus, if there is an equal number of electrons and holes the
thermopower goes to zero, and the overall sign is decided by the
majority carriers.
We find that the thermopower due to the screened charge impurities shows
the same behavior as shown  by the dashed line in Fig.~\ref{fig.2}.
In order to explicitly verify the electron-hole puddle picture near the
charge neutral point we use the effective medium theory, EMT for
graphene \cite{rossi2008b}. 
The density profile close to the Dirac point is quantitatively
described by the Thomas-Fermi-Dirac (TFD) theory \cite{rossi2008a}.
Using the TFD results the transport properties of graphene
close to the Dirac point can be accurately calculated using the EMT.
Denoting by angle brackets disordered averaged quantities,
for the diagonal transport coefficients %, $L_{11}$, $L_{22}$,
from the EMT in 2D we have that the {\em effective medium} coefficients,
$L^{ii}_{\rm eff}$, are implicitly given by the equation \cite{emt_old}:
\begin{equation}
 \left\langle\frac{L^{ii}(\rr) - L^{ii}_{\rm eff}}{L^{ii}(\rr) +
     L^{ii}_{\rm eff}}\right\rangle = 0. 
 \label{eq:emt01}
\end{equation}
Adapting to 2D thermopower the results presented in \cite{webman1977},
the {\em effective medium} off-diagonal coefficient $L^{12}$ is given by:
\begin{align}
  &L^{12}_{\rm eff} = -2 L^{11}_{\rm eff} L^{22}_{\rm eff}
                     \left\langle\frac{L^{12}(\rr)}{(L^{11}(\rr) +
                         L^{11}_{\rm eff})(L^{22}(\rr) + L^{22}_{\rm
                           eff})}\right\rangle\times \nonumber \\ 
  &\hspace{0.15cm}\left\langle\frac{L^{11}(\rr)L^{22}_{\rm eff} +
      L^{11}_{\rm eff}L^{22}(\rr)+ L^{11}_{\rm eff}L^{22}_{\rm
        eff}-L^{11}(\rr)L^{22}(\rr)} 
{(L^{11}(\rr)+L^{11}_{\rm eff})(L^{22}(\rr)+L^{22}_{\rm eff})}\right\rangle^{-1}
 \label{eq:emt02}
\end{align}
%
%Using Eq.~\ceq{eq:emt01} and \ceq{eq:emt02} and the probability
%distribution given by the TFD theory 
%we can calculate the transport coefficients, and in particular the
%effective medium thermopower  
As shown in \cite{rossi2008b} for the local values of $L^{ij}$
we can use $L^{ij}(n(\rr))$ and then using Eqs.~\ceq{eq:emt01},
\ceq{eq:emt02} and the probability distribution given by the TFD theory
we can calculate the EMT transport coefficients, and in particular the
effective medium thermopower
$Q_{\rm eff} =  L^{12}_{\rm eff}/L^{11}_{\rm eff}$ for graphene at,
and away from, the Dirac point.
The results for $Q$ as a function of $n$ at $T=300$~K
for $r_s=0.8$, $d=1$~nm and $n_{i}=10^{12}{\rm cm}^{-2}$
are shown by the solid lines  in Fig.~\ref{fig.2};
in blue (red) are the results
obtained with (without) the effect of the exchange term on the
density distribution. we note that the two-component
model is excellent in describing the main features of the realistic
EMT $Q(n)$ 
close to the Dirac point. 

In conclusion we have developed a complete theory for 
the diffusive thermopower of 2D graphene.  
Quantitative agreement between our theory and existing
graphene experimental thermopower data is a strong indication that the
dominant carrier scattering mechanism operational in 2D
graphene monolayers is screened Coulomb scattering by random charged impurities
located in the graphene environment.
At high densities the Mott formula applies well to the measured thermopower
because of the high Fermi temperature, but it fails in low density limit.
We explain the sign change of the thermopower in the low density regime 
by using both a two component model and a realistic effective medium
theory, that correctly describes 
transport in the presence of the strong carrier density inhomogeneities,
that characterize the graphene density landscape 
close to the Dirac point.
We make a number of specific predictions for graphene thermopower
(e.g. nonlinearity in temperature, existence of a saturation
thermopower at low densities, nontrivial dependence on the background
dielectric constant and on the impurity location), which should be
tested experimentally in order to conclusively settle the issue of
dominant carrier scattering mechanism in graphene.

%\section*{ACKNOWLEDGMENTS}
This work is supported by the U.S.-ONR and NSF-NRI-SWAN.


\begin{thebibliography}{99}


%ER
\bibitem{Kim} Yuri M. Zuev {\it et al.}, arXiv:0812.1393.

\bibitem{Wei} Peng Wei {\it et al.}, arXiv:0812.1411.

\bibitem{Ong} Joseph G. Checkelsky and N. P. Ong, arXiv:0812.2866.

%\bibitem{Kim} Yuri M. Zuev {\it et al.}, arXiv:0812.1393;
%              Peng Wei {\it et al.}, arXiv:0812.1411;
%	      Joseph G. Checkelsky and N. P. Ong, arXiv:0812.2866.


\bibitem{Balandin} A.A. Balandin {\it et al.}, Nano Letters
{\bf 8}, 902 (2008). 

\bibitem{Mott} M. Cutler and N. F. Mott, Phys. Rev. {\bf 181}, 1336 (1969).

\bibitem{thermo_th} T. L\"{o}fwander and M. Fogelstr\"{o}m,
  Phys. Rev. B {\bf 76}, 193401 (2007);
T. Stauber {\it et al.}, Phys. Rev. B {\bf 76}, 205423
(2007);  B. Dora and P. Thalmeier,
Phys. Rev. B {\bf 76}, 035402 (2007); M.S. Foster and I.L. Aleiner,
Phys. Rev. B {\bf 77}, 195413 (2008); M. M\"{u}ller {\it et al.},
Phys. Rev. B {\bf 78}, 115406 (2008).



\bibitem{Ashcroft} N. W. Ashcroft and N. D. Mermin, {\it Solid State
    Physics}, (Thomson Learning Inc. USA, 1976).

\bibitem{Shayegan} L. Moldovan {\it et al.} \prl {\bf 85}, 4369
  (2000).
\bibitem{Lyo} S. K. Lyo, \prb {\bf 70}, 153301 (2004).

\bibitem{Hwang_PRL} E. H. Hwang {\it et al.} \prl {\bf
    98}, 186806 (2007); S. Adam {\it et al}.,
  Proc. Natl. Acad. Sci. USA {\bf 104}, 18392 (2007). 

\bibitem{Hwang_sc} E. H. Hwang and S. Das Sarma, arXiv:0811.1212; \prb
  {\bf 75}, 205418 (2007). 

\bibitem{Hwang_ph} E. H. Hwang and S. Das Sarma, \prb {\bf 77}, 115449
  (2008). 

\bibitem{yacobi} J.~Martins {\em et al.}, Nature Phys. {\bf 4}, 144 (2008).

\bibitem{rossi2008a} E.~Rossi and S.~Das~Sarma \prl {\bf 101}, 166803 (2008).

\bibitem{rossi2008b} E.~Rossi {\em et al.}, arXiv:0809.1425 (2008);
                     M.~Fogler, arXiv:0810.1755 (2008).

\bibitem{emt_old} D.~A.~G.~Bruggeman Ann. Physik {\bf 416}, 636 (1935);
                  R.~Landauer J. Appl. Phys. {\bf 23}, 779 (1952).

\bibitem{webman1977} I.~Webman {\em et al.} \prb {\bf 16}, 2959 (1977).

\end{thebibliography}
\end{document}